\documentclass[12pt]{article}

\usepackage[usenames]{color}
\usepackage{url}
\usepackage[colorlinks = true,
            linkcolor = blue,
            urlcolor  = blue,
            citecolor = blue,
            anchorcolor = blue]{hyperref}

\usepackage[utf8]{inputenc}
\usepackage{charter}
\usepackage{fullpage}
\usepackage{hyperref}
\usepackage{graphicx}
\usepackage{gensymb}
\usepackage{lipsum}
\usepackage[sort&compress,numbers]{natbib}
\usepackage{hyperref}
\hypersetup{hidelinks}
\usepackage[total={6.5in,9in},top=1in,headsep=0.1in,headheight=1in]{geometry}
\usepackage{enumitem,amssymb}
\usepackage[usenames]{xcolor}
\usepackage{xfrac}
\usepackage{bm}
\usepackage{array}
\usepackage{comment}
\usepackage{makecell}
\usepackage{ulem}
\usepackage{siunitx}
\usepackage{xspace}
\usepackage[font={footnotesize,sl}]{caption}
\usepackage{wrapfig}

\newcommand\titletext{Large-Format, Transmission-Line-Coupled \\ Kinetic Inductance Detector Arrays for HEP at Millimeter Wavelengths}

\newcommand\snowmass{
\begin{center}
  \rule[-0.2in]{\hsize}{0.01in}\\
  \rule{\hsize}{0.01in}\\
  \vskip 0.1in
  Submitted to the Proceedings of the US Community Study\\
  on the Future of Particle Physics (Snowmass 2021)\\
  \rule{\hsize}{0.01in}\\
  \rule[+0.2in]{\hsize}{0.01in}\\[-2em]
\end{center}
}

\usepackage[firstpage=true]{background}
\backgroundsetup{contents={\parbox{6.5in}{\snowmass}}, scale=1,placement=top,opacity=1,color=black,position={3.25in,1.2in}}

\usepackage{fancyhdr}
\fancypagestyle{plain}{%
  \fancyhf{}%
  \fancyhead[C]{}
  \fancyfoot[C]{\thepage}
}

\fancypagestyle{empty}{%
  \fancyhf{}%
  \fancyhead[C]{{\it \titletext{} }}
  \fancyfoot[C]{\thepage}
}
\usepackage{acronym}
\acrodef{CMB}{cosmic microwave background}

\acrodef{Al}{aluminium}
\acrodef{AlMn}{aluminium manganese}
\acrodef{TiN}{titanium nitride}

\acrodef{OMT}{ortho-mode transducer}
\acrodef{IDC}{inter-digital capacitor}
\acrodef{JPA}{Josephson parametric amplifier}
\acrodefplural{JPA}{Josephson parametric amplifiers}
\acrodef{KID}{kinetic inductance detector}
\acrodefplural{KID}{kinetic inductance detectors}
\acrodef{TES}{transition edge sensor}
\acrodefplural{TES}{transition edge sensors}

\acrodefplural{KID}{kinetic inductance detectors}
\acrodef{NEP}{noise equivalent power}

\acrodef{TLS}{two-level system}
\acrodefplural{TLS}{two-level systems}

\acrodef{TWPA}{traveling wave parametric amplifier}
\acrodefplural{TWPA}{traveling wave parametric amplifiers}

\newcommand{\wrthz}{W\,Hz$^{-1/2}$}

\def\arcsec{^{\,\prime\prime}}

\def\arcsectxt{$\arcsec$}

\def\vpec{v_{pec}}
\def\vpectxtnosp{$\vpec$}
\def\vpectxt{\vpectxtnosp\ }

\def\mus{\mu\mathrm{s}}

\def\mum{\mu\mathrm{m}}
\def\muev{\mu\mathrm{eV}}
\def\deg{^\circ}
\def\sqdeg{\Box^\circ}

\def\mustxt{$\mus$}

\def\mumtxt{$\mum$}
\def\muevtxt{$\muev$}
\def\degtxt{$\deg$}

\def\permumcu{$\mu$m$^{-3}$}

\def\tinx{TiN$_x$}
\def\sio2{SiO$_2$}

\def\asih{a-Si:H}

\def\nmux{N_{\textrm{mux}}}
\def\Nqp{N_{qp}}
\def\nqp{n_{qp}}

\def\tqp{\tau_{qp}}

\def\s21{S_\text{\tiny 21}}

\def\fr{f_r}
\def\f0{f_0}
\def\qi{Q_i}

\def\qc{Q_c}
\def\qr{Q_r}

\title{\titletext{}}

\date{\today}

\usepackage{authblk}

\author[1]{Peter S. Barry}
\author[1,2]{Clarence. C. Chang}
\author[4]{Sunil Golwala}
\author[2]{Erik Shirokoff}

\affil[1]{Argonne National Laboratory, Lemont, IL, USA}
\affil[2]{University of Chicago, Chicago, IL, USA}
\affil[3]{California Institute of Technology, Pasadena, CA, USA}

\begin{document}

\maketitle

\begin{abstract}
\noindent The kinetic inductance detector (KID) is a versatile and scalable detector technology with a wide range of applications. These superconducting detectors offer significant advantages: simple and robust fabrication, intrinsic multiplexing that will allow thousands of detectors to be read out with a single microwave line, and simple and low cost room temperature electronics. These strengths make KIDs especially attractive for HEP science via mm-wave cosmological studies. Examples of these potential cosmological observations include studying cosmic acceleration (Dark Energy) through measurements of the kinetic Sunyaev-Zeldovich effect, precision cosmology through ultra-deep measurements of small-scale CMB anisotropy, and mm-wave spectroscopy to map out the distribution of cosmological structure at the largest scales and highest redshifts. The principal technical challenge for these kinds of projects is the successful deployment of large-scale high-density focal planes---a need that can be addressed by KID technology. In this paper, we present an overview of microstrip-coupled KIDs for use in mm-wave observations and outline the research and development needed to advance this class of technology and enable these upcoming large-scale experiments.

\end{abstract}

\section{Introduction}

Over the last decade, mm-wave cosmological observations have emerged as a powerful tool for constraining fundamental HEP phenomena. Central to this development has been the advancement of key superconducting mm-wave detector technologies. These developments have enabled ever larger CMB experiments including the upcoming `Stage 4’ cosmic microwave background experiment (CMB-S4). Over the next decade, continued advancement of superconducting mm-wave detector technology will enable even more sensitive instruments that will advance our cosmological understanding in new directions by transforming multiple observables from the mm-wave sky (beyond just the CMB) into precision cosmological probes. The kinetic Sunyaev-Zeldovich (kSZ) effect is already being used to constrain galaxy and cluster peculiar velocities, which can probe cosmic acceleration and test for deviations from GR. Measurement of CMB polarization on sub-arcminute scales would enable future tests on small scales for dark matter self-interactions, for new light thermal particles, and for axion dark matter.  Integral-field spectroscopy at mm wavelengths is an new and emerging capability that can provide an important complement to optical/IR multi-object spectroscopy for measuring both baryon acoustic oscillations (BAO) and redshift space distortions (RSD) via large-scale structure, in particular extending out to much higher redshifts to potentially vastly increase the number of spatial modes accessible.

The realization of the future potential of mm-wave cosmology will require focal plane arrays of detectors with O($10^6$--$10^7$) detectors operating at the fundamental noise limit across the entire mm-wavelength range (30--420 GHz). The 2019 report of the DOE Basic Research Needs Study on High Energy Physics Detector Research and Development~\cite{BRN_HEP_DetRD} identified the need to carry out detector R\&D to achieve this goal (see Section 3.4 and PRD 7,8, and 26 in the report). Flexibility in the optical coupling is a key technical driver, and requires an architecture capable of delivering multi-band photometric and polarimetric imaging for CMB and SZ mapping as well as integral-field spectroscopy for future mm-wave tomographic and line-intensity mapping and spectroscopic surveys. Transmission-line coupled detectors offer a versatile solution for optical coupling by being amenable to many coherent optical reception architectures and spectral selection and even spectroscopy. Simultaneously, kinetic inductance detectors are a massively multiplexable detector technology that can reach fundamental sensitivity limits. The goal of this white paper is to bring together and discuss recent progress toward a flexible, microstrip-coupled architecture for Kinetic Inductance Detectors (KIDs), enabling these important mm-wave cosmological probes of HEP.

\section{HEP Opportunities}

\subsection{Peculiar velocities via the kinetic Sunyaev-Zeldovich Effect}

Observation of the kSZ effect is a new, competitive method to measure the cosmological velocity field to constrain the dark energy equation-of-state and test for modifications of GR. kSZ measurements directly probe the peculiar velocities (\vpectxtnosp) of large objects, while the other method of using velocity measurements, redshift-space distortions (RSD), indirectly probes \vpectxt with smaller objects~\cite{huterer2015a}. RSD and kSZ measurements have different systematics, and so they can complement each other via cross-checks, breaking of parameter degeneracies, and reduction of uncertainties.

One developed method for measuring the small kSZ \vpectxt signal in the presence of noise and the spectrally degenerate CMB primary anisotropy, demonstrated by ACT~\cite{hand2012a}, SPT~\cite{soergel2016a}, and Planck~\cite{ade2016a}, uses correlations between mm-wave maps and O/IR galaxy spectroscopic surveys to statistically detect the pairwise relative velocity between concentrations of mass.  The CMB-HD concept~\cite{cmb_hd_snowmass2021,cmb_hd_website} seeks to exploit this technique via cross-correlation of kSZ and O/IR galaxy survey.  Such a survey would yield an expected uncertainty on the primordial non-Gaussianity parameter $\sigma(f_{NL}) = 0.26$, sufficient to distinguish between multi-field ($|f_{NL}| > 1$) and single-field ($|f_{NL}| \sim 10^{-2}$) models of inflation and roughly a factor of 2 lower in uncertainty than the upcoming SPHEREx survey.

Another approach to kSZ \vpectxtnosp, measuring \textit{individual} cluster peculiar velocities with high precision, could have substantial potential for cosmology. A survey of 30,000 galaxy clusters with $\sigma \textrm{v} = 200$ km/s would yield a Dark Energy Task Force figure-of-merit (FoM) of 170~\cite{bhattacharya2008a}. Combining with Stage IV surveys such as DESI (RSD; FoM 700) and LSST (weak lensing; FoM 800) would improve their FoMs by a factor of two~\cite{mueller2015a}. The same survey would yield a constraint on the cosmological growth index $\gamma$ of $\sigma \gamma = 0.02$~\cite{kosowsky2009a}, sufficient to distinguish modified gravity models such as DGP~\cite{dvali2000a} from GR ($\gamma_{DGP} - \gamma_{GR}$ = 0.13) and complementing comparably precise measurements by LSST, DESI, WFIRST, and Euclid.

To do so requires X-ray temperature information and mm-wave measurements in multiple spectral bands, isolating the kSZ from other, much larger signals: the thermal SZ effect, radio and dusty galaxies, and CMB primary anisotropy. To date, this has been done for fast-moving galaxy cluster substructures with a 10m telescope~\cite{sayers2013a,sayers2019a} pointing to future potential for cluster peculiar velocities. The required combination of sensitivity, angular resolution, and spectral information necessitates a large ($>$30m) mm-wave telescope with a focal plane array covering 6 spectral bands from 90 to 420 GHz. The tens of $\$$M cost of such a large telescope warrants full use of its focal plane, motivating the development of focal plane arrays providing this wide spectral coverage for each spatial pixel. The factor of 4 range in diffraction spot size necessitates preserving the wave nature of the incoming light after optical reception (by, e.g., feedhorns or antennas) so it can be coherently summed in a way to match this varying spot size. The favored tool for this coherent combination is superconducting microstrip transmission line. The technology must scale to ($10^6$) detectors (\# of angular resolution elements in 6 spectral bands for a $1 \degree$ FoV 30m telescope) and must enable spectral bandpass definition.

\subsection{BAO and RSD via mm-wave Integral-Field Spectroscopy}

Another incipient technique at mm-wavelengths for cosmological measurements is integral field, moderate-resolution spectroscopy. This approach combines two features of the mm-wave sky: due to the canceling effects of redshift and the $\nu^{2+\beta}$ behavior of a dusty greybody spectrum, the flux of the dusty/molecular/atomic component of galaxies is roughly redshift-independent at mm wavelengths; and, therefore, the mm-wave sky is dense in sources at even the angular resolution of 30-m telescope; 10" at 300 GHz. Thus, with the FoV for a 30-m telescope at these wave lengths filled with spectrometers, a galaxy will occupy each spatial pixel, and each may be detectable in CO or [CII] spectral line emission. Using these lines to extend BAO and RSD measurements over an enormous range of redshift, $z = 0.5-10$, could be transformational for probing the expansion and cosmological growth function history.

Large format mm-wave spectroscopy will also enable the technique of line intensity mapping (LIM) to be extended to mm-wavelengths. LIM is an observational technique that provides a new probe of large-scale structure using low angular resolution, spectroscopic observations of atomic or molecular line emission to trace the large-scale fluctuations in the matter distribution. Without the need to spatially resolve individual sources, LIM provides access to cosmological modes beyond the redshift reach of traditional optical/IR galaxy surveys by detecting all line-emitting sources in aggregate. Knowledge of the rest-frame wavelength uniquely maps the spectral direction to redshift, leading to a unique dataset comprising maps of the 3D distribution of large-scale  out to high-redshift. Benefitting from the experience of CMB experiments in performing high-sensitivity, low-systematic measurements of faint, diffuse structure at these wavelengths, all that is needed for mm-LIM is the addition of moderate resolution spectrometers in place of existing broadband detectors.

To determine redshifts requires a resolving power $R = \lambda /  \Delta \lambda \sim 300-1000$. The only conceivable approach to provide spectroscopy for each spatial pixel uses superconducting microstrip coupled arrays of mm-wave resonators that transmit narrow spectral bands to individual detectors~\cite{shirokoff2014a}, along with the aforementioned concept for matching to the diffraction spot size. Filling the FoV of a 30m telescope at this $R$ for 90-420 GHz would thus require O($10^7$) detectors.

\subsection{Dark Matter Science from Small-Scale CMB Polarization with CMB-HD}

The CMB-HD concept~\cite{cmb_hd_snowmass2021,cmb_hd_website} seeks to extend the study of CMB polarization to sub-arcminute scales (15\arcsectxt\ resolution) to pursue a range of science:
\begin{itemize}
  \item Use gravitational lensing of the CMB polarization on these scales to measure the matter power spectrum with the goal of distinguishing pure cold dark matter models from those that would possess nontrivial structure on small scales.  Such nontrivial behavior could explain quandaries in the study of small-scale structure such as the paucity of dwarf satellite galaxies and the presence of cores rather than cusps in the density profiles of galaxies.  Such an explanation would be the sign of new dark matter physics, generally some sort of self-interaction.
\item Test for or rule out new light thermal particles at $\ge 95$\% confidence level.
\item Probe for axion dark matter in the cosmos in the
 \muevtxt\ to meV mass range via resonant conversion in galaxy cluster magnetic fields.
\end{itemize}

\section{Technical Requirements}

The technical challenge of the above needs is enormous.  Imaging and polarimetry surveys at sub-arcminute scales will require $\mathcal{O}(10^6)$ detectors over a $\mathcal{O}(10~\sqdeg)$ fields-of-view (FoV) covering 9 spectral bands from 30~GHz to 420~GHz.  Spectroscopic surveys (over a smaller FoV initially, $\mathcal{O}(1~\sqdeg)$, but potentially also reaching $\mathcal{O}(10~\sqdeg)$) will require a further factor of 10--100 increase in detector count.

Importantly, these future focal planes require not only increased detector count, but also increased detector {\it density}. The latter is a physical driver for new technology because existing demonstrated multiplexing schemes used by current experiments~\cite{irwin2002a,hartog2011a,bennett2015a} impose a practical limitation on the detector packing density. For example, the CMB-S4 project uses feedhorn-coupled transition-edge sensor (TES) detectors read out with time-division multiplexing, and the physical size of these multiplexing elements results in a sub-optimal sampling of the high frequency channels. A KID-based architecture eliminates the need for additional cold multiplexing components and is, currently, the only viable path to addressing the eventual ultra-high densities required to meet these new science challenges.

\section{Enabling Technologies}

\begin{figure}[ht]
\centering
\includegraphics*[width=3.75in]{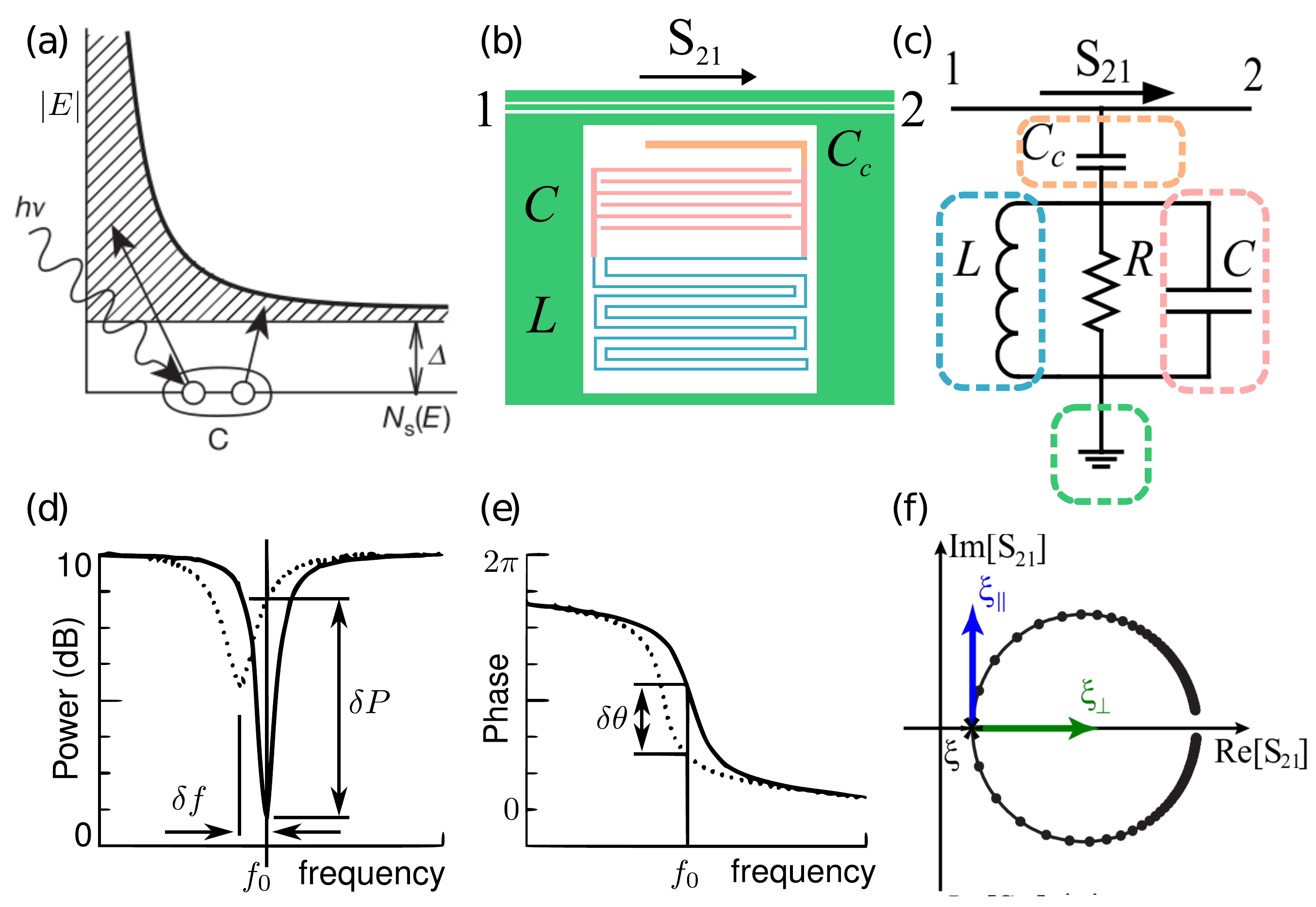}
\caption{Schematic description of \ac{KID} operation. (a) An optical photon breaks a Copper pair in a superconductor, generating quasiparticles. (b) A schematic of a lumped-element \ac{KID} coupled to a microwave feedline. (c) The equivalent LRC circuit of the resonator. The effective values of $L$ and $R$ depend the number of quasiparticles in the superconductor. (d and e) Optical loading causes the frequency, dissipation power (quality factor), and phase of the resonant circuit to shift from the solid to the dotted line. (f) Complex representation of  the response of the detector. Absorbed power leads to both a change in frequency and quality factor, corresponding to motion tangent and normal to the circle, respectively. Figure adapted from~\cite{Mazin:2004tl,Gao:2011ij}.
\label{fig:kid_fig}}
\end{figure}

The \acf{KID}~\cite{Day:2003hh} is a technology that has gained significant traction in a wide range of applications across experimental astronomy over the last decade~\cite{Maloney:2010hh,Monfardini:2010jda,Shirokoff:2012fx,Swenson:2012dd,Galitzki:2014ek,Catalano:2018jh,CastilloDominguez:2018ex,Wilson:2018fy}. A \ac{KID} is a pair-breaking detector based on a superconducting thin-film microwave resonator (Fig. \ref{fig:kid_fig}), where the relative population of paired (Cooper pairs) and un-paired (quasiparticles) charge carriers govern the total complex conductivity of the superconductor. Photons with energy greater than the Cooper pair binding energy ($2\Delta$) are able to create quasiparticle excitations and modify the conductivity. By lithographically patterning the film into a microwave resonator, this modification is sensed by monitoring the resonant frequency and quality factor of the resonator.

The \ac{KID} offers a number of advantages owing to their relative simplicity. Each detector is formed from a microwave resonator with a unique resonant frequency, enabling a large number of detectors to be readout without the need for additional cryogenic multiplexing components. Furthermore, from a fabrication perspective \acp{KID} can be extremely simple, with a number of successful designs consisting of a single metal deposition, lithography and etch process~\cite{Calvo:2016ct}. While more complex architectures are being explored~\cite{Barry:2018el,Endo:2012uy,Johnson:2016kn}, they remain substantially easier to fabricate that other comparable superconducting detector technologies. This is especially important for applications that require high-yield, repeatable, and robust large-format detector arrays.

Experiments based on arrays of KIDs have been developed across a range of frequencies spanning from x-rays~\cite{Cecil:2012go,Giachero:2018ec} to optical/IR~\cite{Mazin:2012kl,Szypryt:2016ix}, down to mm-wave observations~\cite{Johnson:2016kn,Barry:2018el,CastilloDominguez:2018ex}. The low frequency limit is set by the requirement that the photon energy overcome the binding energy of the superconducting pairs. For a conventional superconductor, the binding energy is proportional to the material dependent $T_c$. Typically, \acp{KID} have been based on thin-film \ac{Al} or sub-stoichiometric \ac{TiN}~\cite{Leduc:2010do} with $T_c \sim \SI{1}{K}$,corresponding to $\nu_g > \SI{74}{GHz}$. At higher optical frequency, where the photon energy is large compared to $2\Delta$, \acp{KID} can be operated in `single-photon mode', where pulses from individual photons are resolved yielding a measure of the energy of each photon~\cite{gao2012a}. At low frequency ($\nu<10$ THz), \ac{KID}s have been operated in `integrated' mode, where the average photon flux generates an elevated equilibrium number of quasiparticles. For ground based applications \acp{KID} now routinely achieve sensitivity limited by the inherent fluctuations in the incident photons~\cite{deVisser:2014iw, Mauskopf:2014fb, Hubmayr:2015be}.

In the following sections we present an brief overview of the wide range of KID-based architectures are being developed by a number of groups for a variety of scientific applications.

\subsection{Direct Absorbing KIDs}

The simplest variant of a KID is formed when the resonator geometry is optimized to act as an impedance-matched absorber to efficiently collect the incoming signal. This configuration enables a straightforward path to a densely packed focal plane layout using either an open array architecture~\cite{calvo2016a}, or a waveguide coupled design that has been favored by recent experiments due to the improved control over beam systematics and optical cross-talk~\cite{castillodominguez2018a,wilson2020a}. The optical bandpass for each detector is defined through standardized quasi-optical filtering schemes~\cite{ade2006a} that limit the detectors to a single observing band, with multi-band imaging requiring complex free-space dichroic/trichroic filters. However, in most cases, the simplicity does allow for a smaller per-detector footprint and higher focal plane packing density, which recovers the loss in mapping speed relative to a multi-band configuration at the expense of increased demands on the readout.

\begin{figure}[h]
    \centering
    \includegraphics[width=0.95\textwidth]{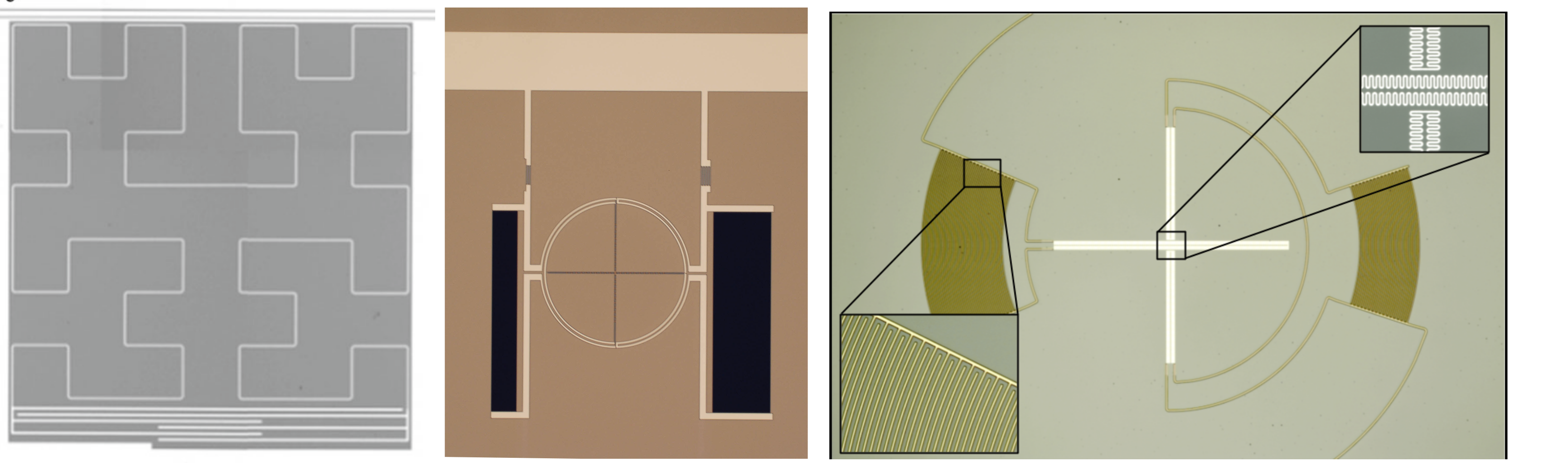}
    \caption{Examples of direct absorbing lumped-element KID architectures. \textbf{Left}) dual-polarisation Hilbert fractal \ac{Al} absorber used by NIKA2~\cite{goupy2016a}, \textbf{middle}) a polarimetric detector design made up of two single-polarisation detectors using \ac{TiN} developed at NIST for Toltec~\cite{austermann2018a}, and \textbf{right}) \ac{Al} for the SPT-4 instrument~\cite{dibert2021a}. }
    \label{fig:directabs}
\end{figure}

To date, the only facility-grade KID-based instruments are based on this detector architecture. With most experiments currently focused on applications in astronomy and local universe astrophysics, the NIKA-2 experiment on the IRAM 30 meter telescope has demonstrated that the KID-based instruments are highly competitive with other approaches. A number of other experiments at major telescope facilities are currently at various stages of development; the Mexico UK Submillimetre Camera for AsTronomy (MUSCAT)~\cite{castillodominguez2018a} and TolTec~\cite{wilson2020a} receivers are undergoing commissioning at the Large Millimeter Telescope in Mexico, the first-light instrument for the Fred Young Submillimeter Telescope (FYST) in Chile, and the next generation receiver for the South Pole Telescope (SPT3G+), are all set to deliver transformational science over the next decade, with FYST and SPT explicitly targeting key HEP science goals through observations of the early and late time SZ effects~\cite{choi2020a,dibert2021a}, as well as Rayleigh scattering of the cosmic microwave background~\cite{beringue2021a}.

\subsection{Microstrip-coupled KID}
Over the past decade, progress in the development superconducting electronics and low-loss transmission lines has opened up a wide range of opportunities for new device architectures and capabilities at mm-wavelengths. The ability to lithographically define circuits capable of on-chip signal processing with extremely low loss has led to major breakthroughs, adopting approaches from the field of microwave/RF engineering, that include delay lines to form superconducting phased-array antennas~\cite{ade2015a}, signal splitters and combiners to construct ultra-compact interferometers, and low/high/band-pass filters as a way of integrating band defining elements onto the same device as the detectors.

The output of any signal processing stage is typically terminated in a power detector. With most implementations now based on superconducting thin-film microstrip transmission lines, a robust approach to couple radiation from the microstrip into a KID with high optical efficiency is needed. Broadly, there are two approaches that are being developed; 1) is based on the proximity coupling of the field to the KID, 2) direct connection of the KID to the transmission line.

\subsubsection{Microstrip Adiabatically Coupled to Parallel-Plate Capacitor KIDs}

\begin{figure}[ht!]
  \begin{center}
    \includegraphics*[width=5.5in,viewport=5 155 790 525]{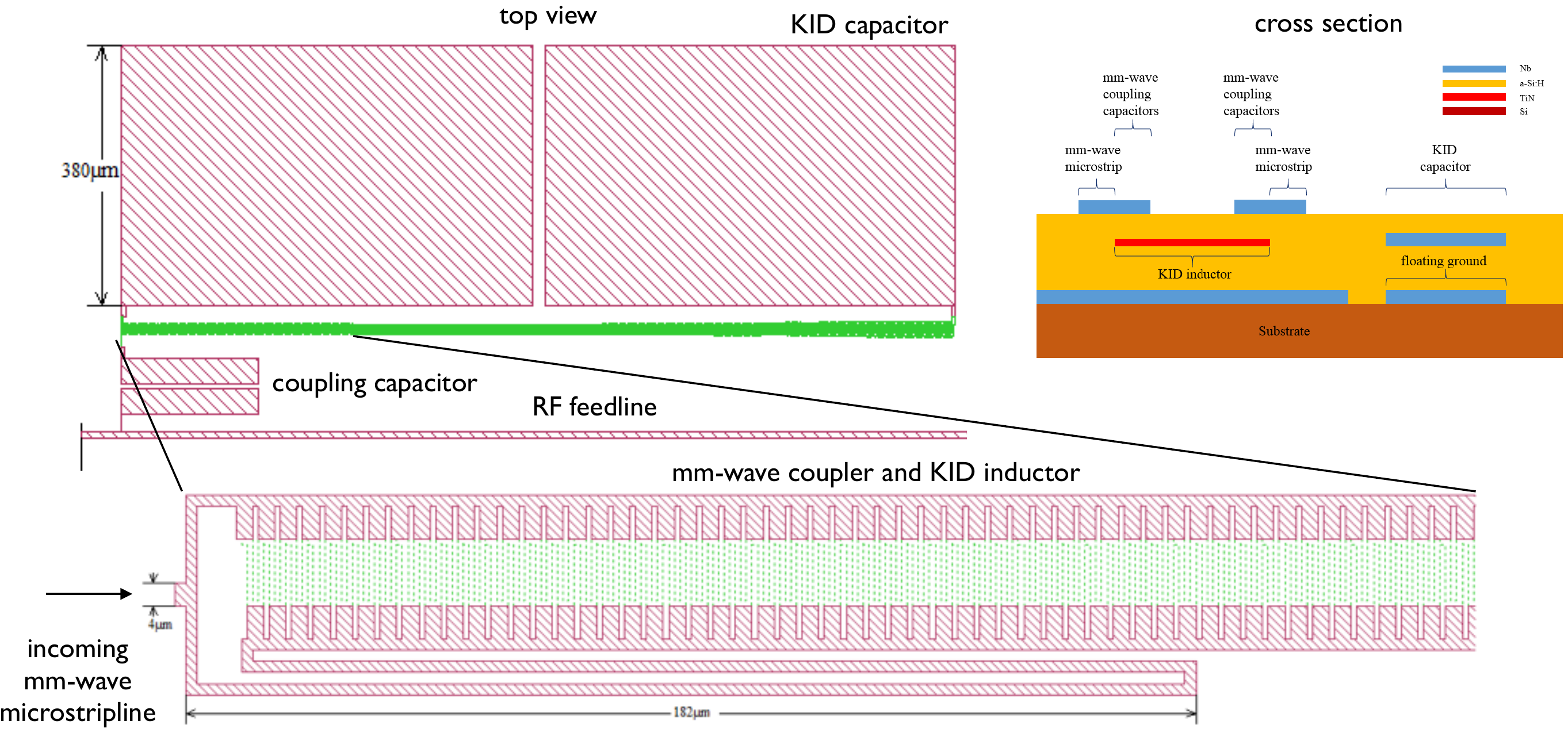}
  \end{center}
  \vspace{-18pt}
  \caption{Microstrip-coupled, \tinx, parallel-plate capacitor KID
    design.  The zoomed-in region shows the incoming microstripline,
    the mm-wave coupler, and the \tinx\ KID inductor (green dashed).
    The microstrip line uses 1100~nm thick \asih\ (two layers,
    800~nm and 300~nm) and 450~nm Nb,
    while the ground plane under the entire structure (not shown) is
    150~nm Nb.  The microstrip line enters a 50-50 splitter, one of
    whose legs undergoes a half-wavelength delay.  The KID inductor
    sits between the two \asih\ layers and meanders between the two
    180\degtxt-out-of-phase microstrip lines, whose top layers have
    extensions overlapping and capacitively coupling to the inductor
    (through the top \asih\ layer).  This coupling drives a mm-wave current through
    the \tinx, which is resistive at these frequencies ($h \nu_{min} =
    2 \Delta$), thereby dissipating that power via Cooper
    pair-breaking.  The weak capacitive coupling is increased
    adiabatically along the microstrip lines so that constant absolute
    power is absorbed per unit length (rather than constant fractional
    power), yielding uniform absorption of essentially 100\% of the
    power over the inductor volume.  The inductor is connected to a
    large, two-element, symmetric, series PPC that uses Nb and the
    bottom \asih\ layer and whose center electrode is an island in the
    ground plane, forming the KID.  The KID is coupled to a RF
    feedline via a similar, smaller PPC. The inductor is about 6~mm long overall, varying by 30--50\% with spectral band, while the
    inductor width and thickness are always 1~\mumtxt\ and 20~nm.  The
    resonant frequencies are 50--100~MHz.  A similar Al design uses 100-nm thick, 1-\mumtxt wide Al and a roughly 30-mm long inductor with resonant frequencies in the 200--400~MHz range.}
  \label{fig:design}
\end{figure}

Figure~\ref{fig:design} illustrates one potential architecture for microstrip-coupled KIDs under development at Caltech/JPL~\cite{multiscale_tin_spie2014, shu_multiscale_al_ltd19}.  The design uses Al or \tinx, the latter tuned for $\Delta = 170$~\muevtxt, corresponding to $\nu_{min} = 80$~GHz and a BCS $T_c = 1.1$~K to enable operation down to the 90~GHz atmospheric window.  Designs are in hand for six bands centered on 90, 150, 220, 270, 350, and 405~GHz, optimal for kSZ (B1 through B6 for brevity).  The design could be matched to \ac{CMB} applications by extending down to 30~GHz (and dropping 405~GHz) by use of lower-$\Delta$ \tinx\ or AlMn and lower operating temperature.

The design has a number of key features that make it unique and flexible.  One is that it uses an adiabatic coupling of the microstrip to the KID to circumvent the high normal-state resistance of \tinx, which, while beneficial for KID responsivity, can be challenging to impedance match to the microstrip line.  A related feature, useful also for the Al design where the impedance mismatch is not a concern, is that the adiabatic coupling provides enough independent design parameters (capacitance between the microstrip and the KID, length of KID meanders, total length of coupler, total KID length, thickness, and width) that it decouples the inductor volume from the mm-wave power absorption optimization.  Another key aspect is that the design uses a parallel-plate capacitor (PPC) with low-noise \asih\ developed at JPL.  A PPC provides inherently low susceptibility to direct absorption by the capacitor of the KID, and it incorporates a ground plane that also shields the inductor, so that light is only incident on the KID via the microstrip in a controlled fashion.  These features make it possible to use this design with a phased-array or lens-coupled antenna in a format that is fully exposed to the incoming light, obviating shielding of the KIDs from direct absorption/stray light (e.g., with a feedhorn array).  In the Caltech/JPL application, a 6:1 bandwidth superconducting phased-array slot-dipole antenna, illuminated through the underlying silicon wafer, feeds the microstrip, which also incorporates bandpass filters (BPFs) to define spectral bands incident on the KIDs.

\paragraph{Performance Results to Date}

\begin{figure}
  \begin{center}
  \includegraphics*[viewport=10 -10 734 500,width=2in]{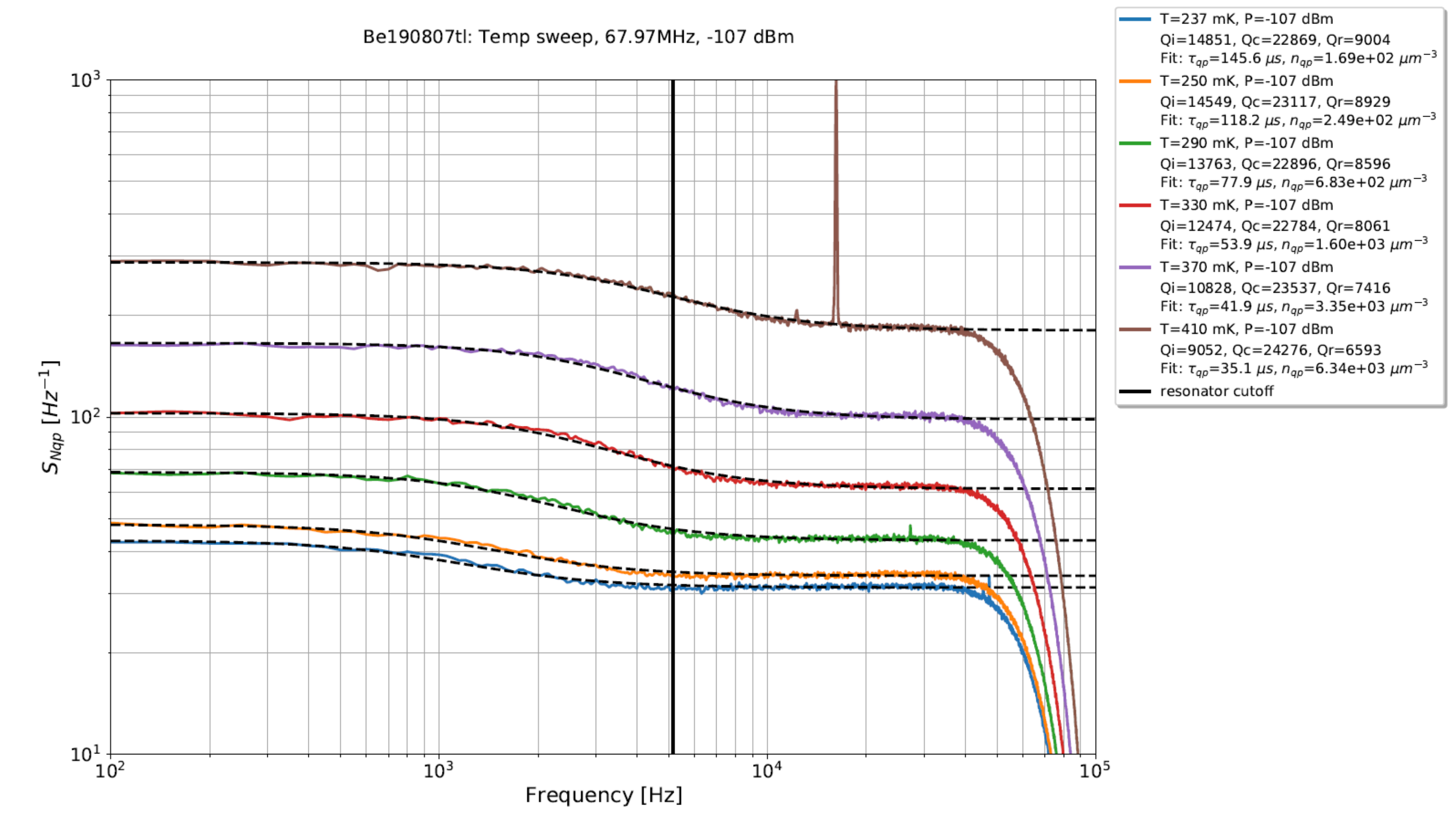}
  \includegraphics*[viewport=0 0 1310 778,width=2.35in]{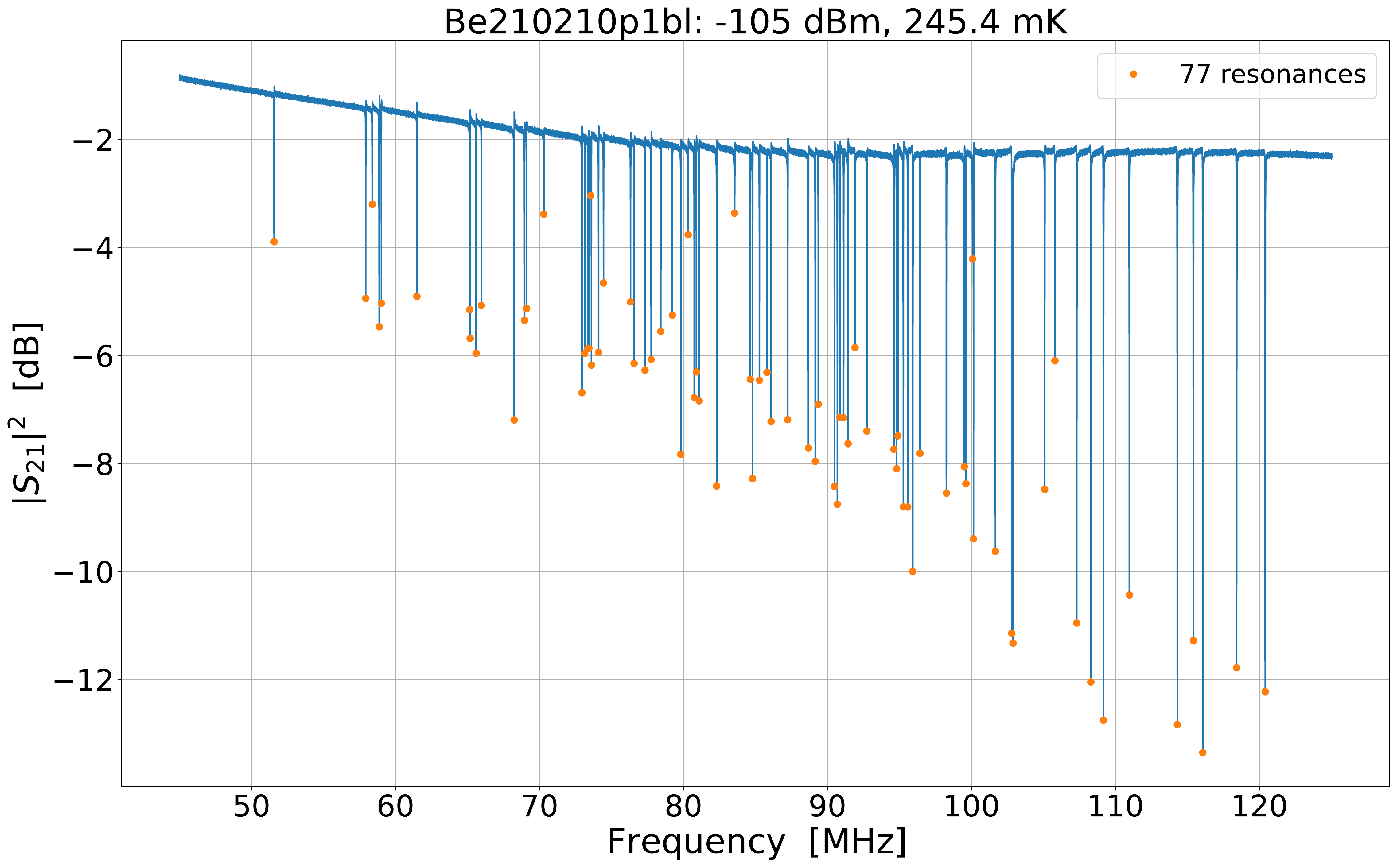}
  \includegraphics*[width=2in]{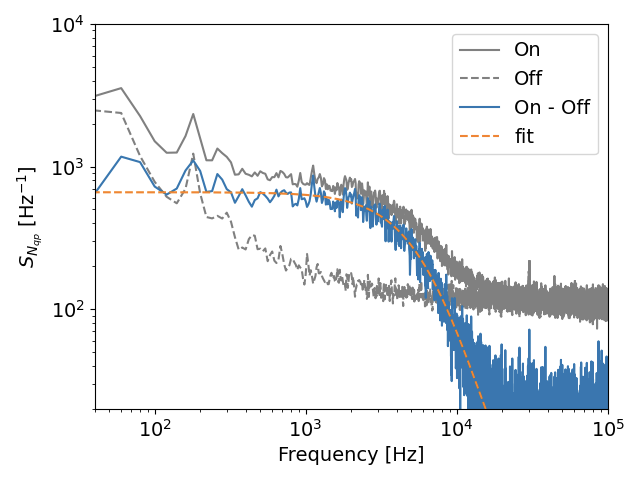}
  \end{center}
  \vspace{-18pt}
  \caption{\textit{(Left)} Thermal generation-recombination noise for a microstrip-coupled \tinx\ KID with $\fr$~=~68~MHz in units of quasiparticles$^2$/Hz, measured under negligible optical load. Data are shown for $T$~=~237, 250, 290, 330, 370, and 410~mK (bottom to top).  The vertical line is the resonator electrical rolloff frequency $\fr/2\qr \equiv 1/(2\pi\tau_r)$.  The resonator coupling quality factor is $\qc = 23$k and the internal quality factor at 237~mK is $\qi = 15$k.  The fits are to a model consisting of the white thermal generation-recombination noise $S_{\Nqp} = 4\,\tau_{qp}\,\Nqp$ rolled off at $f = 1/(2 \pi \tau_{qp})$. The white noise above the rolloff is amplifier noise, which varies with the responsivity.  Lower frequency data are not shown due to insufficiently stable temperature control.
  \textit{(Center)} Transmission spectrum of a high-yield device.  Of 80 expected KIDs, 77 are visible, and the bifurcation powers are about -85~dBm.
  \textit{(Right)} Thermal generation-recombination noise for a microstrip-coupled Al KID with $\fr = 307$~MHz measured at 290~mK under negligible optical load before and after subtraction of low-frequency electronics noise (measured off-resonance).  The fit is to the same model as for the \tinx\ design, though now the amplifier noise contribution is largely subtracted off at high frequency.
  }
  \label{fig:cit_sensitivity}
  \label{fig:cit_yield}
\end{figure}

This design has demonstrated sensitivity sufficient to observe thermal
generation-recombination noise under dark conditions at 240~mK for both the \tinx\ and Al designs.  This is a relevant demonstration because the quasiparticle population under optical load is comparable.  Figure~\ref{fig:cit_sensitivity} shows data acquired under negligible optical load and at a range of temperatures and corresponding thermal values of $\nqp$.  This noise is the stochastic fluctuations in the quasiparticle population due to breaking of Cooper pairs by thermal phonons and the resulting quasiparticles' pairwise recombination.  For various resonators studied, one can infer a thermal $\nqp \approx 150$--500 \permumcu\ for \tinx, $\nqp \approx 2000$ \permumcu\ for Al, and $\tqp \sim 100$~\mustxt\ for both.  This noise implies a limiting noise-equivalent power (NEP) of a few a\wrthz.

\subsubsection{The microstrip-coupled lumped-element kinetic inductance detector}

\begin{figure}[ht!]
    \centering
    \includegraphics[width=0.7\textwidth]{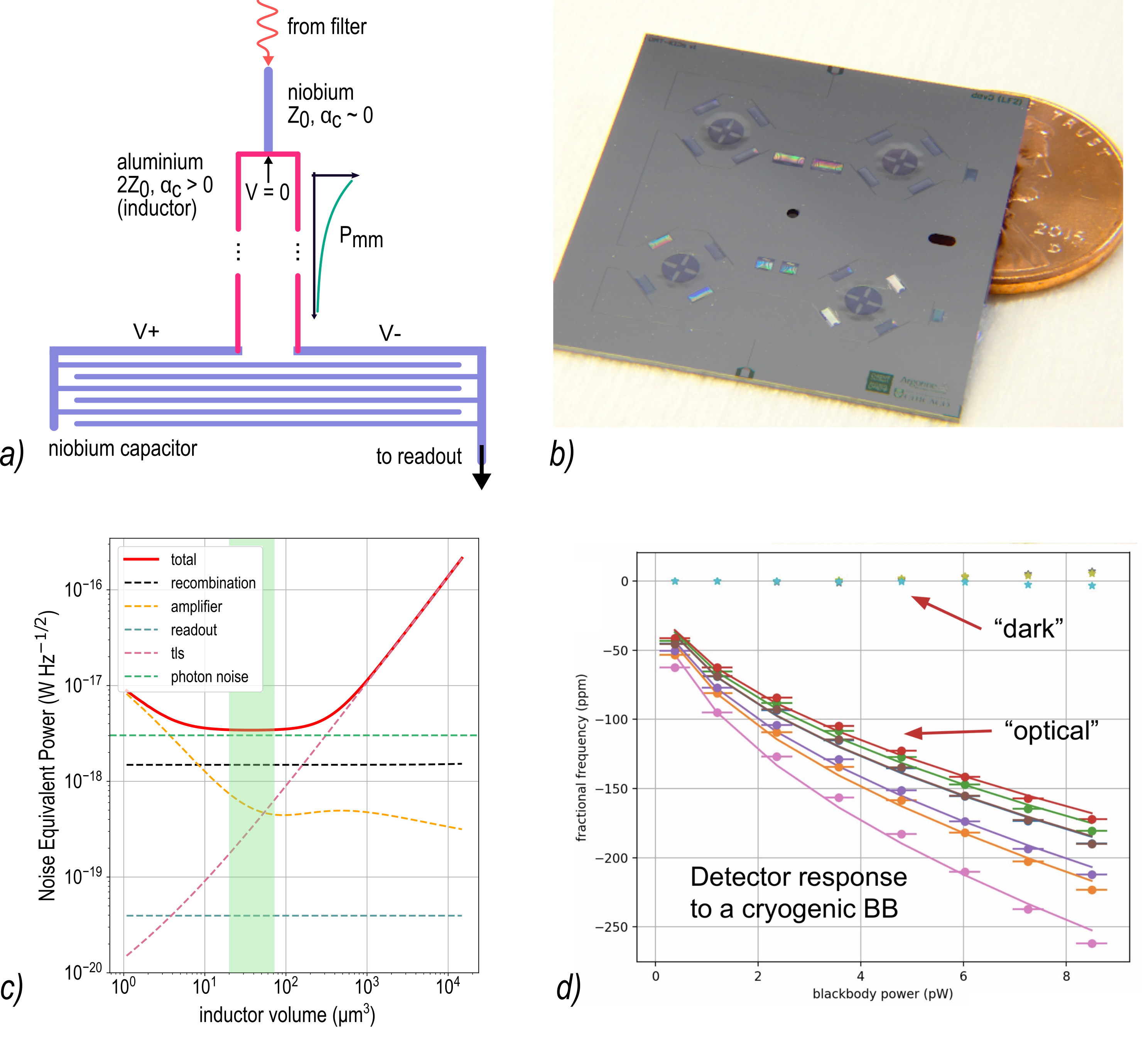}
    \caption{a) Schematic of the principle of operation of a mc-leKID. See text for details. b) Photograph of a prototype mc-leKID device, c) Predicted performance of the mc-leKID as a function of inductor geometry. d) preliminary measurement of optical response characterized with a cryogenic blackbody source as a function of load temperature.}
    \label{fig:mclekid}
\end{figure}

A schematic of an alternative approach being developed at Argonne National Laboratory and the University of Chicago that uses a direct electrical connection is shown in Figure~\ref{fig:mclekid}a, with a photograph of a recent prototype array being developed for a new line intensity mapping experiment for the SPT (SPT-SLIM)~\cite{karkare2022a,barry2021a} shown in Figure~\ref{fig:mclekid}c.

In this design, the superconducting microstrip line feeds the centre (voltage node) of the inductor of the KID resonator, which doubles as a matched, lossy transmission line. As radiation propagates along the KID inductor it is absorbed through the breaking of Cooper pairs and modifies the surface impedance of superconductor. A lumped-element capacitor is used to complete the resonator, and is designed based upon empirical verification of the noise contribution from two-level dielectric fluctuations within the capacitor. The design for SPT-SLIM uses a Nb interdigital capacitor fabricated from the same layer as the mm-wave microstrip circuitry to take advantage of the lower TLS noise expected with the Nb-Si interface~\cite{jonas_arcmp} whilst also minimizing the stray-light cross-section and parasitic inductance.

\paragraph{Performance Results to Date}
One of the primary objectives of the prototype design was to evaluate the suitability of the waveguide-coupled OMT antenna at suppressing parasitic coupling of stray radiation. In previous design iterations that were based on a lens-coupled planar antenna~\cite{barry2018a}, an unacceptably high-level of parasitic optical response was found in the “dark” detectors, which are co-located on the same device, but not connected to an antenna. As well as being an established well understood technology for \ac{CMB} detector arrays, adopting an \ac{OMT} coupling design also provides a high level natural stray-light protection from the metallic horn array, in combination with the membrane coupling that serves as a natural photonic choke that limits the amount of optical power that can enter the bulk-silicon substrate and propagate as surface waves~\cite{yates2018a}.

Figure~\ref{fig:mclekid}d shows the optical response of a set of both optical and dark detectors on the same chip, as a function of incident optical power from a temperature-controlled cryogenic blackbody source. The small level of `dark' response is a major result and highlights the benefit of the \ac{OMT} design, as expected. We then fit the response of the optical detectors to extract an estimate for the optical efficiency for each detector, and observe a spread ranging from $60-70\%$. Whilst promising, the cause of the discrepancy between the measured and simulated efficiency ($>90\%$), as well as the spread is an area of active development. Such a spread could be caused by a misalignment between the waveguide and OMT probes that would modify the coupling efficiency of each horn slightly differently. Investigation into alternative methods of characterizing the Nb-Al microstrip junction are underway to provide an independent validation of the simulation framework.

\subsection{Thermal Kinetic Inductance Detectors}

An alternative method to coupling radiation via a transmission line is to adopt a similar approach that has been developed for the bolometric transition edge sensor arrays.
\begin{wrapfigure}{r}{0.4\textwidth}
\centering
\vskip -12pt
\includegraphics[trim=0in 0.0in 0in 0.0in,width=0.39\textwidth]{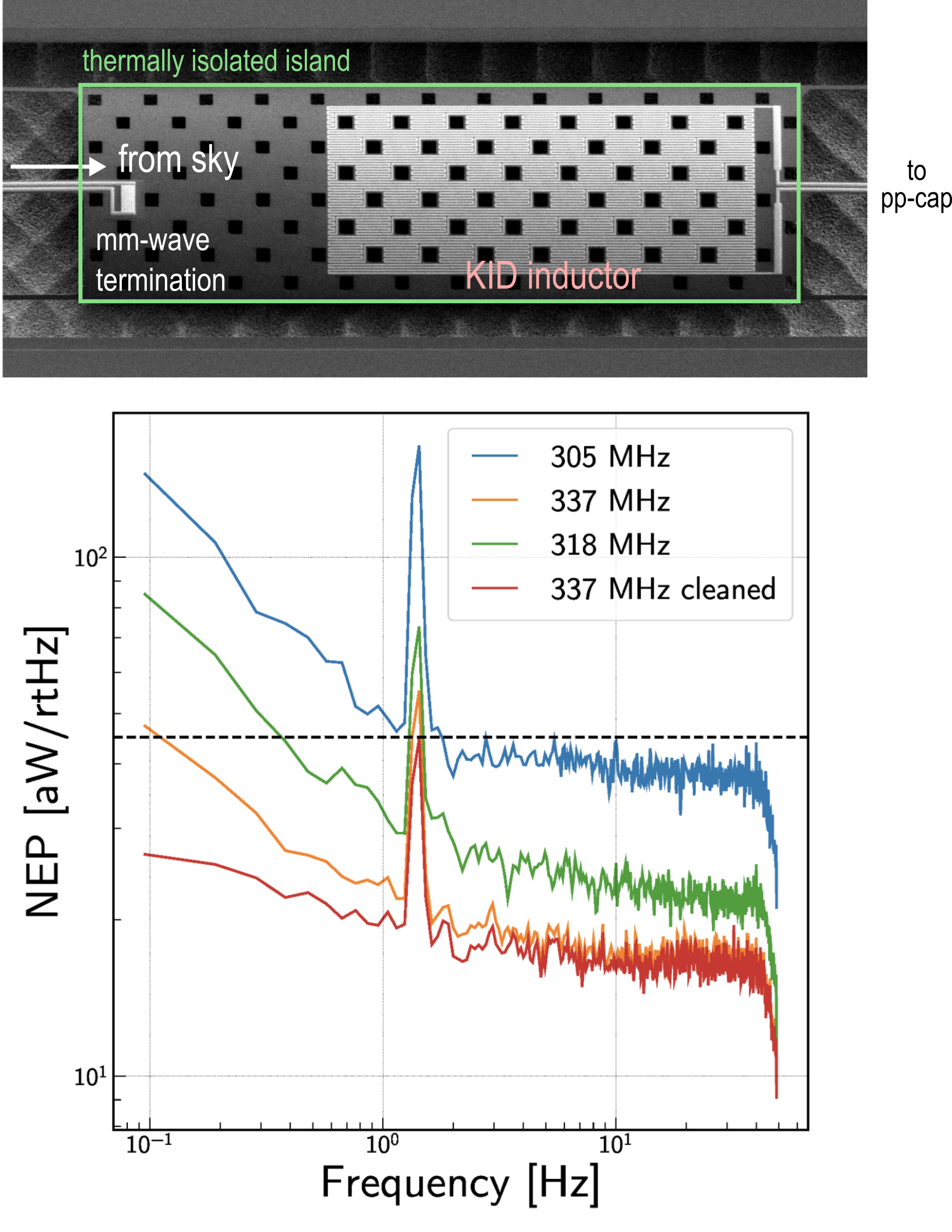}
\vskip -6pt
\caption{mm-wave TKIDs fabricated at JPL. Figure adapted from~\cite{agrawal2021a,wandui2020a}}
\label{fig:tkids}
\vskip -12pt
\end{wrapfigure}
Radiation is absorbed in an impedance matched termination on a thermally isolated island with a KID. Instead of directly absorbed radiation breaking pairs, a thermally-mediated KID (TKID) uses the intrinsic temperature response of the superconducting film to monitor the temperature, and therefore absorbed power.

The primary advantage of the TKID is the ability to independently optimize the resonator geometry and optical coupling, with the latter having undergone extensive development in the context of TES arrays. It combines the multiplexing advantage of KIDs with the proven performance of bolometric designs in TES detectors, at the expense of fabrication complexity. It has recently been shown that the thermal circuit performs similarly to a TES, with the RF readout power providing a mechanism for electrothermal feedback that is essential to achieving the stability and dynamic range required~\cite{agrawal2021a}. Early examples of TKIDs focused on development of sensors for detection of x-rays~\cite{miceli2014a,faverzani2019a}; however, new mm-wave designs for future \ac{CMB} experiments are underway~\cite{steinbach2018a}, as shown in Figure~\ref{fig:tkids}.

\subsection{On-chip spectroscopy}

The natural extension of multi-band imaging using on-chip filters is extension to a filter-bank architecture to realise medium-resolution spectroscopic capability. Several approaches to on-chip spectroscopy exist at a range of technological readiness. Two examples, the recently deployed DESHIMA~\cite{Taniguchi2021} and soon to be deployed SuperSpec~\cite{Karkare2020} instruments employ such a filter-bank design that consists of planar, lithographed superconducting transmission line resonators. Each mm-wave resonator is weakly coupled to the main feedline, and feeds an individual detector. The designs are realized using thin film lithographic structures on a silicon wafer, with the mm-wave circuit formed from high Tc superconducting Nb, NbTiN. For a filter-bank architecture, each pixel in the focal plane is now coupled to $R \approx 100-1000$ detectors, which becomes increasingly challenging for the detector readout. For this reason, in part, the majority of current implementations on-chip spectrometers implement a KID-based backend. However, there are laboratory demonstrations of similar designs constructed using TESs that have been carried out as part of the TIME instrument~\cite{obrient2014a}, as well as for applications in Earth observation~\cite{orlando2020a,goldie2020a} where operating at lower frequency ($<90$ GHz) is beyond the reach of demonstrated materials used for KIDs. Several ongoing projects are working to extend KID designs to both higher and lower frequencies through the use of lower Tc superconducting detectors, higher Tc transmission lines, and free-space cavities as resonators. Very dense focal plane packing may be possible using either parallel plate capacitors with high dielectric constant material such as \asih\ or by modifying the focal plane geometry.

\begin{figure}[ht]
    \centering
    \includegraphics[width=0.75\textwidth]{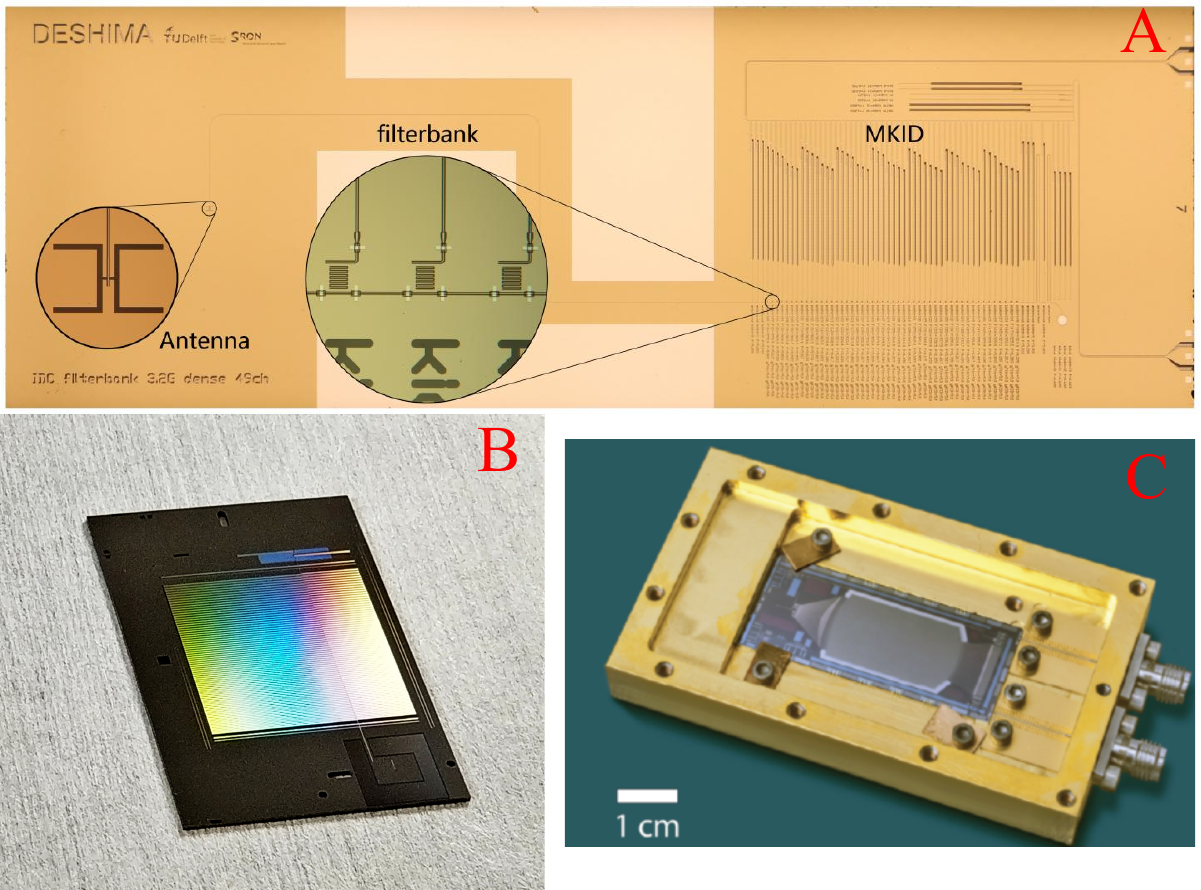}
    \caption{Examples of state of the art on-chip, KID-based spectrometers: (A)DESHIMA (B) SuperSpec (C) Micro-Spec. (see text for references)}
    \label{fig:onchipspecs}

\end{figure}

There are also three alternative techniques currently in development. The first uses multi-path interference within waveguide to effectively create a virtual grating, as demonstrated by the microSpec project~\cite{caltado2019}. The second is the design of an on-chip Fourier transform spectrometer that takes advantage of the current-dependent inductance of a superconducting thin-film to control the wave-speed using an external bias [9]. In this case, the number of detectors per spectrometer is minimized at the expense of needing to sweep the bias appropriately sample the interferogram. The third is the design of a filter-bank spectrometer using free-space transmission elements, either machined in metal blocks as in W-SPEC~\cite{bryan2016a} or fabricated from micro-machined silicon structures.

\section{Future Developments and Path Forward}
Owing to the demonstrated sensitivity and intrinsic ability to scale, KID arrays are now beginning to be adopted by a number of major experiments focusing on astronomy and astrophysics. For future applications in HEP, extending the detection capabilities of KIDs to multi-color/spectroscopic imaging will be transformational, and an overarching goal for KID technology over the next decade will be to develop, validate, and scale these approaches to arrive at a solution, or solutions, to the major technical challenges that are needed for the next generation of HEP experiments at sub-mm/mm wavelengths.

\begin{itemize}
  \item[] \textbf{1. Single pixel optimization: } Development of the existing prototype architectures for large-format multi-band photometers and scalable medium resolution on-chip spectrometers are well underway. The fundamental sensitivity of KIDs has been shown to be comparable to \acp{TES} down to the lowest loading levels~\cite{devisser2014a}, and, when combined with a intrinsic high dynamic range, \acp{KID} are ideally suited to the loading conditions of ground-based HEP experiments.

  The sensitivity under optical load can be estimated for the multi-band photometric detectors using a model based on the data in Figure~\ref{fig:cit_sensitivity}.  Two extremal cases of optical loading characterize the range likely for ground-based instrumentation: under low optical loading as might be expected at 90 GHz on a small-aperture cryogenic telescope, and under higher optical loading as might be expected at 405 GHz on a large-aperture telescope. This model yields total NEPs of 40 and 240 a\wrthz\ for the two cases,  only a factor of 1.8 and 1.4 times larger than the photon noise, and further optimizations can reduce these factors to 1.55 and 1.2.  Use of a lower-gap superconducting material would further reduce the 90 GHz NEP relative to photon noise.\footnote{This is an area of active research, both to enhance sensitivity and to extend the low frequency operating range of \acp{KID}.}  These performances would match that of deployed TES-based instrumentation.  A similar analysis can be performed for the spectrometer pixels using the direct microstrip-coupled architecture, the results of which are shown in Figure~\ref{fig:mclekid}c, indicating that near-photon-noise-limited performance of a \ac{KID} coupled to an $R=300$ spectral channel can be readily achieved.

  Validating these performance projections is an important near-term objective. The intrinsic detector sensitivity and optical efficiency is characterized directly via exposure to a cryogenic blackbody, combined with measurements of the optical bandpass using a Fourier Transform Spectrometer, or a narrow-band high-frequency tunable source. A robust demonstration of near-photon-noise-limited sensitivity with high optical efficiency, long time-scale detector stability, and control over the mm-wave circuitry will build a foundation for larger arrays.

  \item[] \textbf{2. Scaling and cost reduction: } The built-in ability to multiplex large numbers of detectors is a primary   motivation for the \ac{KID}. Achieving high a multiplexing density relies on a combination of inter-related aspects of the detector design, manufacture, and readout: high resonator $\qr$ allows more resonators to be packed into a given readout bandwidth; an optimized level of readout power handling is used to mitigate the noise contribution from the first stage amplifier; and a tuned fabrication process is needed to reduce overlap between resonators (``collisions'') which impacts the overall detector yield. As an example, Figure~\ref{fig:cit_yield} shows no fatal collisions for a \tinx\ device, and two similar devices show non-collided yield $>$95\%.  The Al KID design shows similar yield.  Assuming no improvements, and an expanded $\fr$ range of 50--200~MHz for \tinx\ or 150--600~MHz for Al, a constant fractional frequency spacing matching the current design (0.25~MHz at 85~MHz), and a comparable collision rate of $< 5$\%, we expect the current design is consistent with $\nmux~\approx~500$ at 95\% yield, which is similar to the baseline adopted by several recent KID-based instruments~\cite{gordon2020a,karkare2020a,wilson2020a}. Further improvements offer the potential to reach $\nmux = 1000$--2000 in the near future through a combination of increased control over fabrication processes and employing post-fabrication capacitor trimming~\cite{liu_ctrimming2017,shu_ctrimming2018}, which uses post-testing ion-beam or laser trimming or etching to modify $C$ and thus $\fr$ to fix any collisions.

  As mentioned above, future multi-band focal planes on large field-of-view telescopes will require O($10^6 - 10^7$) detectors in a single focal plane, and a simple scaling of existing implementations is unlikely to be sufficient for realizing the large-format focal plane arrays envisioned for future HEP experiments. For example, a fully sampled spectroscopic focal plane with $R = 1000$ operating over an octave bandwidth on an existing telescope with $3000$ spatial pixels requires $\sim3$M detectors.

  The resonator $\qr$ is set by a combination of the choice of material, detector geometry, and expected optical loading.
  A 10$\times$ improvement in multiplexing over existing implementations, to $10^4$~res/octave, would require $\qr \sim 10^5$ (assuming resonator-to-resonator spacing of 10 linewidths), an optimistic but feasible improvement over currently achieved quality factors in large arrays.  With current readout technology, $\sim300$ octave-bandwidth readout channels would be required, each with its own readout line and cryogenic amplifier, which begins to place impractical demands on the design of the cryogenic subsystem through excessive heat loading.  From a readout perspective, assuming a readout power of $-95$~dBm/resonator (lower than typical), $10^4$ resonators on a single channel will result in $-40$~dBm readout power incident at the cryogenic amplifier (assuming crest factor of 15 dB), which approaches the compression point of current low-noise cryogenic amplifiers.  It will almost certainly be necessary to employ tone-tracking~\cite{dober2021} and carrier-nulling.

  \textbf{Scaling to such large arrays will require dedicated support and a focused programme of development in key technologies to achieve the ambitious technical goals for future generations of HEP experiments.}

  An important goal for scaling up KID technology is reducing the full per detector (i.e. including readout) cost. A long term target is $\sim \$1$/detector, requiring a reduction of 100x from existing experiments (e.g. TolTec $\sim \$150$/det~\cite{wilson2020a,toltec_nsf_website}, CMB-S4 ($\sim \$250$/det)~\cite{abazajian2019a}).

  The cost of the detector production benefits greatly from existing effort and investment in stage 3 and stage 4 \ac{CMB} experiments that have established a foundation of processes and techniques needed for large-scale fabrication of superconducting detector arrays. \ac{KID}-based detectors typically require fewer fabrication steps relative to a standard \ac{TES} process optimized for \ac{CMB}, and addressing the remaining constraints on material quality and process control will leverage this previous infrastructure. Therefore, the focus should be mainly on reducing the complexity and cost of the detector readout where there are number of approaches that promise substantial gains. The continual advances in the semiconductor industry providing new platforms, such as the RF system-on-a-chip, that are under active development offer a route toward high-frequency electronic platforms capable of direct digital sampling of microwave signals up to 6 GHz, ideally suited to the needs of future KID readout. However, while operating a single channel over multiple octaves of bandwidth is a possibility that is being explored~\cite{swenson2012a}, a full-scale implementation is needed to demonstrate that the intermodulation distortion of the signal chain can be adequately controlled as the number of tones is increased. Ideally, a final optimized version of the electronics would be implemented on dedicated application specific integrated chips (ASICs) in order to reduce the volume cost and streamline performance. Furthermore, cryogenic amplifier technologies optimized for low DC dissipation could mitigate many of the practical design challenges, but the implications of a likely lower compression point for channel count would need to be understood. Dissipation-less superconductor-based amplifiers~\cite{eom2012a,bockstiegel2014a} could be an attractive alternative, but would also need focused development on optimization of power handling or would require tone-tracking/carrier-nulling.

  \item[] \textbf{3. Integrated performance: }

  The long term HEP goal for the KID research and development is deploying large experiments to carry out cosmic surveys. As major facilities operating for many years, these surveys constitute significant investments and require technologies that have demonstrated high technical maturity. Achieving the needed technical readiness requires building fully integrated detector systems and operating them in a relevant environment. The only approach is to deploy smaller-scale systems onto existing mm-wave telescope facilities and to operate these instruments to observe the mm-wave sky. These deployments and observing campaigns provide an ideal evaluation of full end-to-end receiver-level performance and establish the foundation needed to develop the scientific approach. As part of the overall technology research and development, it is imperative that these on-sky technology validations are carried out in parallel with the research activities described above.

  A relevant example is the Summertime Line Intensity Mapper for the South Pole Telescope (SPT-SLIM)~\cite{karkare2022a,barry2021a}.  Serving as both a technical and scientific pathfinder, SPT-SLIM leverages the SPT platform to advance line intensity mapping at mm-wavelengths (150 GHz) using on-chip spectrometer technology. Initially targeting the astrophysics through observing the integrated line-emission from nearby galaxies, SPT-SLIM serves as a necessary prerequisite for developing and constraining key aspects of the HEP spectroscopy science case~\cite{karkare2022b}.

\end{itemize}




\pagebreak
\bibliographystyle{unsrturltrunc6}

\bibliography{references,pb}  
%
%
%
\end{document}